\documentclass[10pt]{article}

\usepackage{graphicx}
\usepackage{dcolumn}
\usepackage{bm}
\def  \bea  {\begin{eqnarray}}
\def  \eea  {\end{eqnarray}}
\def  \nn   {\nonumber}
\begin{document}
\begin{center}
{{\LARGE \bf Drag and Diffusion coefficients in extreme scenarios of temperature and chemical potential}} \\
\bigskip
{\large \bf Sreemoyee Sarkar$^*$\footnote{email: sreemoyee.sarkar@saha.ac.in}} \\
{\large \bf Abhee K. Dutt-Mazumder $^*$\footnote{email: abhee.dm@saha.ac.in}} \\
{$^*$Saha Institute of Nuclear Physics, Kolkata, India} 
\bigskip
\end{center}

\begin{abstract}
A comparative study of high and zero temperature plasma for the case of damping rate, drag and diffusion coefficients 
have been presented. In each of these quantities, it is revealed how the magnetic interaction dominates over the
 electric one at zero temperature unlike what happens at high temperature.

\end{abstract}


\section{Introduction}
  The study of nuclear matter under extreme conditions has been an active field of research for the past few decades. 
In particular the 
creation of quark-gluon plasma at high temperature has drawn special interest both in the theoretical and experimental fronts
 in view of 
  different experimental programs like RHIC, LHC etc.
 The other domain of QCD phase diagram where chemical potential is considered to be much higher than the temperature is less
  explored. This is the region of interest for the upcoming experiments
on compressed baryonic matter to be performed at
FAIR/GSI. The high density QCD or QED plasma has also its relevance to astrophysics. In particular it is known that 
the ultradegenerate plasma might exist in the core
 of the neutron stars.  
 
 In plasma, particles propagation get modified through the interaction with the surrounding
medium (quasiparticles). These quasiparticles are the relevant degrees of freedom in terms of which the dynamics of the plasma has to be understood.
 In this work we focus on three quantities {\em viz} the damping rate, drag and momentum diffusion coefficients of the plasma.
 The latter two are intimately connected with the former as exposed in the current work. The study of quasiparticle 
damping rate ($\gamma$)
as well as the transport coefficients like drag and diffusion coefficients ($\eta, {\cal B}$) in case
 of high temperature plasma have been studied for last decades \cite{Iancu97, Thoma91, Moore05}, whereas computation of these
 coefficients in case of degenerate plasma is relatively unexplored field of research. 
Calculation of all these quantities, as is known, suffer from infrared divergences for bare Coulomb or magnetic interaction.
 How
 these divergences are handled in the finite temperature by using the techniques of hard-thermal loop (HTL) 
\cite{Pisarski90} have been well studied.
 We also encounter similar divergences at zero temperature where the hard-dense loop (HDL) \cite{Manuel96} corrected 
propagator
 for the intermediate boson is used to remove such divergences. Here, we discuss the departure of behavior of high temperature plasma from the zero 
temperature one. To reveal
 these differences further and understand the dynamics evolved in each cases we first focus on $\gamma$ subsequently we discuss
 $\eta$ and ${\cal B}$.


\section{Quasiparticle damping rate in hot and dense plasma}
\begin{figure}
\begin{center}
\includegraphics[scale=0.4]{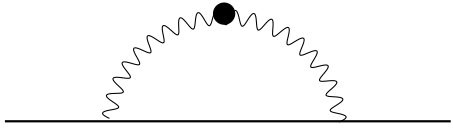}
\caption{Fermion self-energy with resummed photon propagator.}
\label{fig9}
\end{center}
\end{figure} 
The quasiparticle damping rate in the field theoretical language can be written in the terms of self-energy. Assuming the energy
 of the quasiparticle to be hard $E\sim\mu$,
\bea
\gamma\equiv -\frac{1}{4p}\rm{tr}\left[{p\llap{/\kern1pt}}\rm{Im}\Sigma\left(p_0+i\epsilon, {\bf p}\right)\right]|_{p_0=p},
\label{def_gamma}
\eea
 $\Sigma$ is the Fermion self-energy,
\bea
\Sigma (P)=e^2 T\sum_s \int \frac{d^3q}{(2 \pi)^3}\gamma_{\mu}S_0(P-Q)\gamma_{\nu}\Delta_{\mu\nu}(Q).
\eea
$S_0(P-Q)$ and $\Delta_{\mu\nu}(Q)$ are the free fermion propagator and dressed photon propagator respectively. Detail structures of these
 propagators can be found in \cite{Sarkar10}. Explicit evaluation of Eq.(\ref{def_gamma}) gives,
\bea
\gamma&=&{\pi e^{2}\over E} \int{{\rm d}^3q\over
(2\pi)^3}\int_{-\infty}^\infty{{\rm d}k_0\over 2\pi}\rho_f(k_0)
\int_{-\infty}^\infty{{\rm d}q_0\over 2\pi} \nonumber\\
&\times &
(1+ n(q_0)- {\bar n}(k_0))  \delta(E-k_0-q_0)\nonumber \\
&\times &[p_0k_0+{\bf p\cdot
k}+m^2] 
  \rho_{l}(q_0,q) \nonumber \\
&+&2[p_0k_0-({\bf p\cdot \hat q})({\bf k\cdot\hat
q})-m^2]\rho_{t}(q_0,q), 
\label{eqno(8)}
\eea
$\rho_{l,t}$ are the imaginary part of the photon propagator. 
The calculation of the quasiparticle damping rate 
is plagued with infrared divergences in case of soft photon momentum exchange. To remove the divergences one can in principle use the 
Braaten and Yuan's prescription (BY) \cite{BY}. In this prescription an intermediate momentum scale $q^{*}$ is introduced, 
so that for the exchanged momentum region
 $q>q^{*}$ one can use the bare propagator and in the other domain where $q<q^{*}$ HTL/HDL resummed propagator
  has to be incorporated. In case of high temperature relativistic plasma $\gamma$ has been studied earlier in \cite{Iancu97}.
 The phase space factor in the high temperature limit can be written as $(1+ n(q_0)- {\bar n}(k_0))\sim T/q_0$, with this and
 HTL propagator the
quasiparticle damping rate remains divergent in the transverse sector giving \cite{Iancu97},
\bea
\gamma_t&=&\frac{g^2T}{2\pi^2}\int_0^{q^{*}}\frac{dq}{q}\tan^{-1}\left(\frac{3\pi \omega_p^2}{4 q^2}\right),
\eea
while for the longitudinal case we have \cite{Iancu97},
\bea
\gamma_l&=&g^4T^3/m_D^2.
\eea
In the zero temperature limit the phase space factor takes the following form $1+n(q_0)=\theta(q_0)$ and
 ${\bar n}(k_0)=\theta(\mu-E-q_0)$ due to Pauli blocking. In this case ($T=0, \mu\neq0$) one can eventually show that 
with the help of the Pauli blocking and HDL propagator
 finite $\gamma$ can be obtained \cite{Manuel97,Ollitraut97},
 \bea
 \gamma_l&=&\frac{e^2 \nu^2m_D}{64},\nn\\
\gamma_t&=&\frac{e^2\nu m_D}{24\pi}.
 \eea
 Here, $|E-\mu|/m_D=\nu<<1$, which means the scattering of quasiparticles take place very close 
to the Fermi surface and the results are finite.
  \section{Quasiparticle drag and diffusion coefficients in hot and dense plasma}
  The quasiparticle drag coefficient is related to the energy loss of the particle,
  \bea
  \eta=-\frac{1}{E}\frac{dE}{dx}
  \eea
  where, $-dE/dx$ is the particle energy loss which can be obtained by weighting the interaction
   rate with the energy transfer per scattering. Hence, one can obtain the expression for the drag coefficient as follows,
   \bea
-\frac{dE}{dx}=\int d \Gamma \omega,
\eea
$\Gamma$ is the interaction rate. In case of high temperature plasma with bare propagator in $2\rightarrow2$ scattering process
 $\eta$ shows logarithmic divergence ${e^4T^2\over 24\pi E}\int {dq\over q}$ whereas in case of
 ultradegenerate plasma the divergence is even more worse ${e^2 (E-\mu)^3
 m_D^{2}\over 16\pi E}\int{{dq}\over q^4}$. In both the cases BY prescription can be used to show that finite 
results can be obtained \cite{Sarkar10}.
 Even the algebraic divergence of ultradegenerate plasma can be removed with the help of the HDL propagator. We eventually have
 to evaluate the following expression in two extreme cases $T\neq 0, \mu=0$ and $T=0, \mu\neq0$,
\bea
 \Big(-{dE\over dx}\Big)  &\simeq& {e^2 m_D^2  \over 4 \pi }   
 \int  dq_0  dq(1+n(q_0)-\bar n(E-q_0-\mu))  \nonumber\\
  &\times& \Big\{
 {q_0^{2}\over \{2 \left[q^2 + m_D^2 Q_l({q_0\over q}) \right]^2  +
 {m_D^4 \pi^2 q_0^2 \over  2q^2}\}}  
\nonumber \\
 &+&{ q_0^{2}\over\left[ 2q^2 + m_D^2 Q_t({q_0\over q})  \right]^2  +
 {m_D^4 \pi^2 q_0^2 \over 4q^2 }}\Big\},
\label{eloss}
\eea
where,  
\bea
Q_l(x)  =  1 -  \frac{x }{2}
\ln{\frac{1+x} {1- x} } \ ,
\qquad 
Q_t (x)  =  - Q_l (x) + {1 \over 1 - x^2  } \ .
\label{eqno(13)}
\eea
The denominator of the Eq.(\ref{eloss}) can be expanded in powers of $x$ and for the longitudinal exchange the denominator
 can be replaced by $(q^2+m_D^2)^2$ and for the transverse interaction $q^4+(\pi^2m_D^4x^2)/4$,

In the high temperature plasma the phase space factor can be approximated as $T/q_0$ as mentioned earlier, with this we obtain the 
result for $\eta$ \cite{Moore05},
\bea
\eta&\simeq&{e^4T^2\over 36\pi E}
\Big(-{1\over 2}+{3\over 2}\rm{log}\left|{q_{max}\over m_D}\right|
+{1\over 2}\rm{log}\left|{2\over \sqrt \pi}\right|\Big),\nn\\
\label{drag_high_t}
\eea
where, $q_{max}$ can be set by the kinematics of the problem. In the limit where $T=0$ and $\mu\neq 0$ 
the drag coefficient turns out to be \cite{Sarkar10}, 
\bea
 \eta\simeq{e^2 m_D^2\nu^2 \over 48 \pi E}+{e^2 m_D^2\nu^3  \over 96  E}
+O(\nu^4).\nn\\
\label{drag_zero_t}
\eea
In the last expression the first term corresponds to the magnetic interaction and the second term corresponds to the electric one. 
Hence, from the above result it is evident that in case of the zero temperature plasma the longitudinal and transverse modes 
contribute at different order whereas in case of the high temperature plasma there is no such splitting.  

Another transport coefficient which could be of importance to study the equilibration property of the quasiparticle is the 
momentum diffusion coefficient ($B_{ij}$).
 $B_{ij}$ is defined via the relation, 
 \bea
 { B_{ij}}=\int d\Gamma q_i q_j.
\label{diff_def}
\eea
$B_{ij}$ can be decomposed into longitudinal ($B_l $) and transverse components ($B_t$) as follows,
\bea
 B_{ij}=B_t (\delta_{ij}-\frac{p_i p_j}{p^2}) + B_l \frac{p_ip_j}{p^2}.
\label{diff_decom}
\eea
Eventually $B_{l,t}$ can be obtained by weighting the interaction rate with the longitudinal/transverse momentum transfer 
in a collision with the plasma. One can proceed in the same way as that of the drag coefficient to obtain ${\cal B}$ in two
 extreme scenarios ($T\neq 0,\mu=0$ and $T=0, \mu\neq0 $) respectively \cite{Moore05,Sarkar10}, 
\bea 
{\cal B}&\simeq&{e^4T^3\over 18\pi}\Big(-{1\over 2}+{3\over 2}\rm{log}\left|{q_{max}\over m_D}\right|
+{1\over 2}\rm{log}\left|{2\over \sqrt \pi}\right|\Big)\nn\\
{\cal B}&\simeq&{e^2 m_D^3\nu^3 \over 72 \pi }+{e^2 m_D^3 \nu^4  \over 128 }
+O(\nu^5).
\label{diff_coeff}
\eea
From the Eqs.(\ref{drag_high_t}), (\ref{drag_zero_t}) and (\ref{diff_coeff}) it is seen that physics changes 
significantly from high to zero temperature in both $\eta$ and ${\cal B}$. The logarithmic dependence in high temperature 
changes to algebraic form in the ultradegenerate limit. 

\section{Summary}
The main concern of the present work is to reveal the differences of well known high temperature results
 with their zero temperature counterparts. Interestingly it is seen that although the calculation of $\gamma,\eta$ or 
${\cal B}$ involve very similar starting point the final result however differ qualitatively mainly because of the presence
 of the Pauli blocking. For example we find that at zero temperature for the excitations near the Fermi surface leading contributions 
emanate from the transverse sector while
 the Coulomb interaction contribute at the subleading order. The infrared behaviour of the quantities we calculate here
 also exhibit dissimilar infrared behaviour. In fact the usual logarithmic divergence what one encounters become worse with
 algebraic dependence. In the end however they give finite results as we have shown above.


\noindent
\begin{thebibliography}{50}
\medskip
\bibitem{Iancu97} J. P. Blaziot and E. Iancu, Phys. Rev. D {\bf 55}, 973(1997).

\bibitem{Thoma91} E. Braaten and M. H. Thoma, Phys. Rev. D {\bf 44}, 1298(1991).
\bibitem{Moore05} G. D. Moore, D. Teaney, Phys. Rev. C {\bf 71}, 064904(2005).
\bibitem{Pisarski90} E. Braaten and R. D. Pisarski, Nucl. Phys. B {\bf 337}, 569
(1990).
\bibitem{Manuel96}  C. Manuel, Phys. Rev. D {\bf 53}, 5866(1996).

\bibitem{Sarkar10} S. Sarkar and A. K. Dutt-Mazumder, Phys. Rev. D {\bf 82}, 056003(2010).
\bibitem{BY} E. Braaten and T. C. Yuan, Phys. Rev. Lett. {\bf 66}, 2183(1991).
\bibitem{Manuel97}  M. Le Bellac and C. Manuel, Phys. Rev. D {\bf 55}, 3215(1997).

\bibitem{Ollitraut97} B. Vanderheyden and J. Ollitrault, Phys. Rev. D {\bf 56}, 5108(1997).

\end{thebibliography}
\end{document}